# Exhaled Breath Analysis for Monitoring Response to Treatment in Advanced Lung Cancer

Inbar Nardi Agmon MD[a],[†] Manal Abud BSc[b, †], Ori Liran MSc[a], Naomi Gai-Mor MSc[a], Maya Ilouze PhD[d], Amir Onn MD[c], Jair Bar MD PhD[c], Rossie Navon BA, RN[c], Dekel Shlomi MD[a], Hossam Haick PhD[c,*] and Nir Peled MD PhD FCCP[a, d,*]

[a] The Thoracic Cancer Research and Detection Center, Sheba Medical Center, Tel-Aviv 52621, Israel.
[b] The Department of Chemical Engineering and Russell Berrie Nanotechnology Institute, Technion – Israel Institute of Technology, Haifa 32000, Israel.
[c] Thoracic Oncology Unit, Institute of Oncology, Sheba Medical Center, Tel-Aviv 52621, Israel
[d] Thoracic Cancer Unit, Davidoff Cancer Center, Rabin Medical Center, Israel

[†] Inbar Nardi Agmon and Manal Abud contributed equally to this study.

* Corresponding authors:

Hossam Haick
*The Department of Chemical Engineering and Russell Berrie Nanotechnology Institute, Technion – Israel Institute of Technology, Haifa 32000, Israel.*
Tel: +972 (4) 8293087
Fax: +972 (4) 8295672
Email: hhossam@technion.ac.il

Nir Peled
*Thoracic Cancer Unit, Davidoff Cancer Center,
RMC, Kaplan St, Petach Tiqwa, Israel  49100*
Tel: +972 (0)3 9377958
Tel (Fax): +972 (0)3-9378146 (8047)
Email: nirp@post.tau.ac.il

**Introduction:** RECIST criteria serve as the gold standard method to monitor treatment efficacy in lung cancer. However, time-intervals between consecutive CT scans might be too long to allow early identification of treatment failure. This study examines the use of breath sampling to monitor responses to anti-cancer treatments in patients with advanced lung cancer.

**Methods:** 143 breath samples were collected from 39 patients with advanced lung cancer. The exhaled breath signature, determined by GC-MS[1] and nanomaterial-based array of sensors, was correlated with the response to therapy assessed by RECIST: Complete Response (CR), Partial Response (PR), Stable Disease (SD), or Progressive Disease (PD).

**Results:** GC-MS[1] analysis identified 3 volatile organic compounds (VOCs) as significantly indicating disease control (PR/SD), with one of them also significantly discriminating PR/SD from PD. The nano-array had the ability to monitor changes in tumor response across therapy, also indicating any lack of further response to therapy. Using one sensor analysis, 59% of the follow-up samples were identified correctly. There was 85% success in monitoring disease control (SD/PR).

**Conclusion:** Breath analysis, using mainly the nano-array, may serve as a surrogate marker for the response to systemic therapy in lung cancer. As a monitoring tool, it can provide the oncologist with a quick bed-side method of identifying a lack of response to an anti-cancer treatment. This may allow quicker recognition than the current RECIST analysis. Early recognition of treatment failure could improve patient care.

---

[1] gas-chromatography/mass-spectrometry

**INTRODUCTION**

Monitoring treatment efficacy by non-imaging bedside methods is crucial in lung cancer, where The Response Evaluation Criteria in Solid Tumors (RECIST) assessment is performed every 2-3 cycles and sometimes even less frequently. A non-imaging method is particularly important when disease may respond to treatment, although the target lesions increase in size (pseudo-progression), as sometimes occurs with immunotherapies. Therefore, understanding treatment efficacy and its failure early in the course of therapy is extremely important and may improve treatment outcome [1-3]. To date, imaging has been primarily used in the assessment of response to anti-cancer treatment in lung cancer patients, Computerized Tomography (CT) being the mainstay. The RECIST [1-4] approach categorizes the total response into one of 4 groups: complete response (CR), partial response (PR), stable disease (SD) or progressive disease (PD; see 'Methods') [4]. However, time-intervals between consecutive CT scans might be too long to allow early identification of treatment failure; moreover, scans are expensive and not always available.

An emerging approach that shows great promise in monitoring lung cancer (LC) treatment is based on volatile organic compounds (VOCs) emitted in breath samples [5-14]. These VOCs emanate from the membrane of the cancer cells and/or from its surrounding tissues to the blood stream as a result of inflammation and oxidative stress, being excreted by diffusion across the pulmonary alveolar membrane and exhaled through the breath [15-17]. The role of VOCs as markers in various lung diseases, including cancer, has already been evaluated. Analysis of exhaled breath has already successfully distinguished asthmatic patients from healthy individuals [18, 19], delineated COPD [20, 21], differentiated between patients with LC and healthy patients with non-cancerous lung nodules [5, 6, 11, 14, 16, 22-27] , and between subtypes of lung cancer [28]. Broza et al [29] have shown that breath analysis can be used for short-term follow-up after LC-resection. However, changes in VOCs in LC patients throughout the course of their disease and under systemic anti-cancer therapy have yet to be studied.

This study investigated the possibility of breath VOC signature being a marker for treatment efficacy in advanced lung cancer patients. Exhaled breath analysis is a simple non-invasive method that could allow the oncologist to recognize disease status during the course of treatment. In this prospective study, consecutive breath samples were collected from patients receiving systemic therapy for advanced stage LC. The samples were categorized into different response groups by matching them to their evaluation from sequential CT scans of the response of tumors to treatment. Several comparisons were conducted to identify differences in VOC patterns between the response groups. Two separate breath-analysis methods were used for this purpose - gas chromatography with mass spectrometry (GC-MS) and chemical nano-arrays (NA-NOSE). While GC-MS allowed the identification and quantification of a wide variety of separate breath VOCs, the NA-NOSE provided a sensor-based discrimination between the groups without regard to the nature and composition of the breath VOCs.

## PATIENTS AND METHODS

### Patients

Between May 2012 and June 2013, 39 patients diagnosed with stage III/IV lung cancer and scheduled for systemic therapy (chemotherapy or targeted therapy) at the Thoracic Oncology Unit of the Sheba Medical Center, Israel, were enrolled. Approval was obtained from the ethics committee of the Sheba Medical Center [NCT01386203]. All patients were given a written and verbal explanation about the study prior to their enrollment, and had signed a written consent of participation form. Each patient was interviewed regarding personal data, smoking habits and relevant medical history. Clinical characteristics of the study population are summarized in Table 1.

### Breath collection

Exhaled alveolar breath samples were collected from patients in a controlled manner. The first breath sample was taken from each patient prior to and close to the start of the first systemic therapy cycle. Repeat samples were collected prior to and adjacent to subsequent cycles, or during follow-up meetings with the attending physician.

Patients were asked to refrain from smoking, and withhold alcohol and food consumption for 2 h prior to breath sampling. Alveolar exhaled breath was collected in chemically inert Mylar bags (Eco Medics, Duernten, Switzerland) in a controlled way after a 1 min procedure of a lung washout procedure described elsewhere [10, 11, 23]. Breath collection was a single-step process that did not require changing between the dead space and alveolar breath bags. The content of each bag was transferred immediately through an offline procedure to a coded Tenax® sorbent tube (SKC Inc., Eighty Four, PA). The tubes were stored at 4ºC in a clean environment and transported to the Laboratory for Nanomaterial-Based Devices (Technion, Israel) for analysis by both GC-MS and the sensors' nano-array.

(**Figure 1**)

### Methods

**Chemical analysis of breath samples**

GC-MS were used to identify the exact composition and seek specific informative breath VOCs [7, 71, 87]. Gas chromatography uses a helium stream that carries the sample through a long heated capillary column to separate molecules in the VOCs mixture according to their volatility, more volatile compounds travelling faster than less volatile ones. Due to the difference in the chemical properties in the mixture, each molecule has a different exit time from the column (the retention time). The mass spectrometer determines molecular mass and chemical structure of the VOCs after being broken up into characteristic fragments and ionized. In the mass analyzer, the ions are filtered by an electric field according to their mass/charge (m/z) ratio. The retention time and the mass charge ratio are used to identify the compounds by a spectral library in the GC software, thus reducing the possibility of error in identification. Certain identification could be done only by the analysis of pure standards to validate the supposed identity by the system [88]. (For more details, see Supplementary Information (SI), section 1.2)

**Nanoarray Analysis of Breath Samples**

The nanomaterial-based sensor array used to analyze the breath samples contained 40 cross-reactive, chemically diverse chemiresistors that were based on 2 types of nanomaterials: (i) organically stabilized spherical gold nanoparticles (GNPs, core diameter 3-4 nm), and (ii) single wall carbon nanotubes (SWCNTs) capped with polycyclic aromatic hydrocarbon (PAH). The chemical diversity of the sensors was achieved through different organic functionalities. The GNP and SWCNT/PAH or SWCNT/HBC sensors responded rapidly and reversibly when exposed to typical VOCs in the breath [24, 25]. We also confirmed that they are only weakly responsive to water [23, 30, 31]. (For further details, see SI, section 1.3)

**Study Design**

**Evaluation of patients' response to treatment**

Patients went through periodical CT scans to evaluate their response to treatment. Referral to scans was decided upon by the attending physician, regardless of the study. CT scans were reviewed centrally by an expert radiologist. The patient's response was categorized from each CT scan relative to the previous scan [26] as:

    (i) **Complete response (CR)** - disappearance of all target lesions.

(ii) **Partial Response (PR)** – a 30% decrease in the sum of the longest diameters of target lesions.
(iii) **Progressive Disease (PD)** – a 20% increase in the sum of the longest diameters of target lesions.
(iv) **Stable Disease (SD)** - small changes not meeting other criteria.

The first breath sample of each patient was taken as the **Baseline Sample (BL)**. Subsequent samples were matched with the next proximal CT scan of the patient, and categorized according to one of the 4 abovementioned groups. Analysis included only breath samples that met one or more of the following criteria: (1) a breath sample that was taken up to 3 months prior to the last available CT scan; (2) a breath sample that was taken up to 3 weeks after the last available CT scan. If more than one CT scan was available that fulfilled the criteria, the sample was categorized by the latest scan. Three main groups were defined for comparison after categorizing the breath samples by response:

(i) **Baseline samples**
(ii) **Disease Control (DC)** - composed of PR+SD samples, which both indicated a positive response to treatment (i.e. no worsening of the disease)
(iii) **PD samples**

The clinical status of each event was defined relative to the previous disease status.

**Statistical analysis**

The breath samples were analyzed using 2 independent approaches in 2 phases. In the first phase, a GC-MS system operating with a post-run analysis program (GCMS solutions version 2.53SU1, Shimadzu Corporation, Germany) was used to find statistically different concentrations between the 4 groups. Compounds were tentatively identified through spectral library matching (Compounds Lbrary of the National Institute of Standards and Technology, Gaithersburg, MD 20899-1070, USA). In the second phase, breath samples were characterized with an array of nanomaterial-based sensors, combined with multivariate discriminant function analysis (DFA) [34] as statistical pattern recognition algorithms (SI, section 1.4), with the aim of identifying the "breath print" of the different defined groups. The accuracy of DFA was confirmed by employing leave-one-out cross validation.

The Wilcoxon test was used to find statistically significant differences in the average abundance of the VOC's for the different groups (SI, section 1.5). Data classification was attempted by Discriminant Factor Analysis (DFA). The accuracy of DFA was further confirmed by leave-one-out cross-validation, and sensitivity, specificity and accuracy were calculated. Statistical analysis was carried out using SAS JMP, Version 10.0 (SAS Institute Inc., Cary, NC, USA, 1989-2005).

## RESULTS:

A summary of the clinical data of each of the advanced LC patients who participated are given in **Table 1.** Approximately 80% were males, mean age 62 ± 7.2 years. Former smokers accounted for 69% (average 52 ± 24 pack years); 18% were current smokers (average 86 ± 54 pack years), and 13% had never been smokers (**Table 1**). Most (85%) of the patients had Non-Small Cell Lung Cancer (NSCLC) with different histological diagnoses (squamous cell, adenocarcinoma and "not otherwise specified" (NOS)). Only 15% of the patients had Small Cell Lung Cancer (SCLC). (**Table 1**). Stage 3 patients had also been treated with radiotherapy, although outside the study period. About 82% of the patients received chemotherapy as first-line therapy, the most common protocol being a combination of Cisplatin/Carboplatin and Pemetrexed (Alimta). About 20% of the patients were treated by EGFR/ALK TKIs (EGFR: 7 ALK: 1).

Breath samples (143 in total) from 39 patients were collected, ranging from 1 to 12 per patient (average 3.7). A total of 135 samples meeting the inclusion criteria were analyzed. The average time between subsequent breath samples was 6.7 (± 6.3) weeks. The first sample was taken on 03/15/2012, and the last sample on 06/27/2013. The average time of follow-up was 18.1 (±17.7) weeks (range 0.1-64 weeks).

BL numbered 39 samples, while there were 47 samples in the PR group, 38 in the SD group, and 11 in the PD group (**Table 2).**

**Chemical analysis of the breath samples**

In the first phase, we identified VOCs that could be useful as biomarkers in monitoring the response of a cancer to treatment. By GC-MS analysis, ~200 relevant VOCs were identified, all with main masses in the range of 33 to 282 m/z and retention times between 1.6 and 39.8 min. Seventeen of these VOCs were found in >90% of the samples, with a similarity index >85%. Using these 17 VOCs, binary comparisons were conducted, as given above in detail under 'Methods', and summarized in **Table 2**:

(i) **DC (PR/SD) events after one cycle of therapy: Baseline samples vs. consecutive samples (i.e. first samples taken under treatment) of DC (PR + SD):** 27 patients.

(ii) **PD events after one cycle of therapy: Baseline samples vs. consecutive samples of PD:** 5 patients progressed immediately at the first sampling time. These were too few for a reliable comparison.

(iii) **PR vs. SD:** 47 and 38 SD samples, respectively.

(iv)     **DC vs. PD:** 85 PR+SD samples and 11 PD samples.

The non-parametric Wilcoxon/Kruskal-Wallis test with a cutoff value of P=0.05 was used to find the significant VOCs. Three VOCs were identified as significant markers for DC for the first breath sample taken under treatment (**Table 3**), namely **alpha-phellandrene**, **styrene** and **dodecane-4-methyl**. **Styrene** discriminated significantly between samples categorized as DC compared with PD. GC-MS analysis failed to show any significant discriminant VOCs between PR and SD cases.

**Nano-array analysis of the breath samples:**

In the second phase of the study, 131 breath samples were exposed to nano-array sensors, of which 4 samples were excluded for technical reasons. Thirty-eight were BL samples, 48 were PR, 34 were SD, and 11 were PD.

We examined whether the sensor array could estimate and monitor tumor responsiveness to treatment, thereby alerting us about possible changes in tumor behavior. For this purpose, we developed 2 predictive models based on breath samples collected from 38 patients using DFA. Twenty-seven patients presented SD or PR at the first assessment, in whom the first DFA model attempted to identify disease control (DC samples) after the first treatment compared with the BL. This model showed a sensitivity of 93%, a specificity of 85% and an accuracy of 89% after leave-one-out cross-validation. The Positive Predicted Value (PPV) and Negative Predicted Value (NPV) for identification of DC were 86 and 92%, respectively.

In the second DFA model, we distinguished 83 samples categorized as DC from 11 samples categorized as PD. Despite achieving 100% specificity and 92% classification accuracy, the sensitivity was only 28%, whereas the PPV and NPV for identification of PD were 100 and 91%, respectively. Because of too few numbers of samples, we did not analyze patients who had PD in the first few cycles.

Our main challenge was to monitor the single patient and be alerted to any change in the response to therapy, which could be achieved by using HBC-3 sensor. **Figure 2** is a hot-plot in which each patient is represented in a separate row, and repeated breath samples are shown in separate columns. Each value has been normalized to the previous categorized sample. SD values for each categorization, set by calculating means and standard errors, were: (i) PD >0.15, (ii) -0.05 ≤ SD ≤ 0.05 and (iii) PR<-0.05 (all units are given in ohm).

The hot-plot represents single-sensor analysis of 127 breath samples from 33 patients. Thirty-three of the samples were BL, and 94 were follow-up samples (49 PR, 34 SD, 11 PD). Using one sensor, breath analysis correctly identified 59% of the PR (29/49), 59% of the SD (20/34) and 54% of PD (6/11) samples. Altogether, 59% (55/94) of the follow-up samples were correctly identified. For Disease Control, i.e. SD and PR together, there was an 85% success in response monitoring. There was no difference between different histological categories in this result.

**DISCUSSION:**

The term "tumor response assessment" was introduced long ago in the medical literature. Miller et al. [31] as early as 1981 discussed the importance of creating a 'common language' to be used in evaluating response to anti-cancer treatment. Numerous investigations of tumor response assessment have since been carried out in different types of cancer, including lung cancer [35]. Recent progress in cancer therapy has raised this challenge again, since response to immunotherapy and targeted therapy may now have to be judged on a different scale, as yet undefined.

The World Health Organization (WHO) recommends response criteria for solid tumors, such as lung cancer, and proposes the use of the computed tomographic (CT) scan and endoscopy as methods to classify tumor response [36]. Along with the CT scan as the mainstay, advanced and novel imaging techniques are used to provide better information about tumor volume, perfusion and glucose uptake as potential biomarkers for tumor response and clinical outcome [35].

Nevertheless, although imaging remains the main method for objectively monitoring tumor response to treatment, it has some major disadvantages. The high cost and limited availability of imaging modalities, such as CT and CT/PET, preclude their use at short intervals for assessing patients' responses to treatment. According to the WHO recommendations, "progression in any site indicated disease progressions, despite objective responses in other sites" [36].

Protein biomarkers, including CEA, CA125, CYFRA 21, NSE and ProGRP, can also be used in monitoring disease status [37], but this is only relevant when tumor markers are increased. Taking all these facts into consideration, a need for additional methods for monitoring tumor response to treatment becomes apparent, especially ones that enable closer follow-up and are more up-to-date in the bid for more reliable clinical decision making.

This study indicates that exhaled-breath analysis may serve as a simple, user-friendly method of monitoring response to systemic therapy in lung cancer. Its particular importance is its potential to recognize failure very early on in the course of therapy before it becomes manifest by routine imaging schedules. This may help the physician to make better, more up-to-date, rational decisions regarding the cancer treatment selected for each patient. Likewise, in cases in remission (e.g. case #17 in Table 8 and Figure 5), exhaled breath may serve as a surrogate marker of disease recurrence.

Most importantly, both methodologies could recognize disease control within the first assessment, which is particularly important as it might avoid inefficient therapy that might have significant adverse effects.

Conceptually, consecutive breath-samples would be collected from a patient at fixed times after receiving a cycle of anti-cancer treatment before initiating another cycle of therapy. Because of its simplicity, breath-collection can be done the physician's office. By using the signals received from the sensor-array, the physician could create a patient's personal 'hot-plot'(as in Figure 2). Its scale allows the physician to assess the patient's response to treatment and take appropriate clinical action.

In this study, 2 powerful tools were used: (i) GC-MS. which is effective in the identification and quantification of human volatolomics. Gas chromatography separates VOCs according to their volatility, and the mass spectrometer determines their molecular mass and possible chemical structure. This method has some considerable limitations that preclude its wider use in clinical practice, including: expense, the need for pre-concentration techniques, a high level of expertise in its operation and analysis, and its time consuming nature. (ii) **Nano-arrays**. This approach uses a cross-reactive sensors array, each responding differently to all or part of the chemical compounds in an exhaled breath sample. By applying pattern recognition algorithms and classification techniques, the combined responses of the sensors can establish odor-specific response patterns without actually separating and identifying the specific VOCs responsible for causing the odor [38, 39].

**Use of volatile biomarkers for monitoring response to treatment**
The source of the exhaled VOCs is not well understood [40]. We identified 3 VOCs as markers for response (Table 3). Several explanations have been published regarding their formation.
**Styrene** is an aromatic compound used in manufacturing of plastics, synthetic rubber and resins that is considered to be an exogenous pollutant. Other exogenous sources include tobacco smoke, alcohol, other types of pollution and radiation. The IARC classifies styrene as a possible carcinogen to humans [41]. Since aromatic compounds can be highly reactive, they can be taken up into the cytoplasm, attacking tissues and organs in the body by causing peroxidation damage to proteins, polyunsaturated fatty acids (PUFA) and DNA. This damage accumulates during the lifetime, possibly leading to age-dependent diseases, such as cancer, and products from it may be released slowly into the breath [42] .

**a-Phellandrene** (a-PA; 5-isopropyl-2-methyl-1,3-cyclohexadiene) is a cyclic unsaturated monoterpene formed from essential oils of many edible plants or herbs.

**Dodecane, 4-methyl** is a branched-chain alkane.

The main mechanism affecting the emission of hydrocarbons in the body is oxidative stress, which is defined as the overall balance between formation and scavenging of reactive oxygen species (ROS) and free radicals in the body. ROS are molecules or ions with unpaired electrons in the outer shell, which are constantly being produced in the mitochondria as part of cellular respiration. They can also be produced by fungal and viral infections, or stem from exogenous sources, such as cigarette smoke, pollution and radiation [40] .

The hypothesis that various VOCs have different origins in the body could explain the different changes in concentration between the VOCs in response to anti-cancer treatment. For example, some VOCs can be produced locally within the tumor or its microenvironment; their concentration VOCs would be expected to drop if there is a positive response to treatment and a reduction in tumor volume, but to rise with disease progression. Other VOCs might be produced as a result of systemic changes induced by presence of a cancer, and changes in concentration would therefore be expected to take longer to be expressed in the breath. On the other hand, some VOCs might be released from tumors cells undergoing necrosis, such that their concentration might rise in the case of a positive response to treatment.

Taking everything into consideration, identification of specific VOCs that change in concentration as a result of changes in the cancerous disease may point the goal towards more precise, focused research in the future, hopefully with larger sample size. **Styrene** appears to be the most interesting VOC in our study, being the only one that showed significant changes in concentration in both comparisons (Baseline samples vs. DC, and DC vs. PD). In both cases, the concentration of styrene was lower in the DC group, suggesting that **a decrease in the concentration of styrene in exhaled breath can indicate a positive response to treatment.** This tendency of a lower concentration of styrene associated with improvement in the cancerous disease has already been noted [29, 40]. As most of our cohort responded during the first few cycles, our upfront PD group was too small to investigate the level of this VOC compared with baseline samples, an interesting issue requiring futher investigation.

**Use of sensor-array for evaluation of treatment response**

The sensor-array analysis provides an easier and more intuitive indication. Analysis achieved a relatively high accuracy in the 2 sets of classifications. The first DFA model (see 'Methods') took into consideration patients who had reached disease control after the initial treatment (i.e. SD or PR). Their baseline samples compared with their subsequent samples showed a relatively high classification capability (93% sensitivity and 85% specificity). The second DFA model compared PR/SD samples with subsequent PD status; the specificity (true negative value) in this model was 100%, but the sensitivity was only 28%, indicating that it is highly reliable in detecting negative progression of the disease and lack of response to treatment (PD states).

In practice, these results indicate that the physician can predict a positive response to treatment starting from the first session with a PPV of 86%. Following initial assessment, the physician can use sensor set 2 to test whether the patient had a positive response (PR/SD) or did not respond to treatment, i.e. the disease had progressed (PD). In this case, the PPV was 100%, implying that identification of a sample as PD is 100% trustworthy. These are only preliminary results that require a larger study group for validation; nevertheless, the results support the potential use of the NA-NOSE as a monitoring tool and show its ability to discriminate between response and lack of response to anti-cancer treatment.

The attempt to use only one sensor in monitoring the response to treatment led to an interesting pattern presented as a hot-plot in Figure 2, which displays each patient separately. Each sample is normalized to the previous sample from the same patient, and is colored according to the resistance of the sensor for the specific sample. If one assumes correct identification in achieving **Disease Control** (SD+PR) *vs.* **lack of response to treatment** (PD), **85% percent of the cases were identified and colored correctly** by the single sensor (HBC-3). There may be several explanations regarding the cases that were wrongly identified. First, some cases could have been categorized wrongly, mainly because sometimes the true volume or progression of the cancer is not well represented in the CT scans - either the changes might still be too small to trace, or a new tumor focus exists in a part of the body that was not scanned and therefore was not recognized. In such cases, samples categorized as PR/SD might actually be PD. In the same way, a sample categorized as SD might actually be PR if a positive response to treatment has been detected by the sensor, but not as yet by the CT scan. A second possibility is that, in this hot-plot, each sample is normalized to the previous one; but in some cases, the time-interval between the 2 samples has been too long (due to clinical reasons, e.g. prolonged lack of evidence of disease), and normalization then gives an inaccurate result.

**Future potential and clinical applications**

GC-MS has many disadvantages such as being time-consuming, costly, requiring cumbersome equipment and specialized knowledge in order to analyze the results. Therefore, the nano-array method is a more realistic option for use in clinical practice in the future, requiring only compact and easy to use equipment, and does not require pre-concentration and/or dehumidification procedures. The sensors are broadly cross-reactive, respond to a wide range of compounds, and are less affected by noise than the detected (sub) ppb-ppm concentrations of the compounds identified by the GC-MS. Moreover, the sensors can be tuned to show very low sensitivity to confounding VOCs from the surroundings, whereas the GC-MS detects these confounding VOCs, thereby introducing noise into the measurement of the compounds of interest, which is likely to affect the overall accuracy of the method [5, 6, 16, 29] .

The limitations of this study include a relatively small cohort group. However, we had a total of 143 comparisons and an internal longitudinal monitoring was also used for all the patients, with a longest period of 14 months.  We enrolled patients under range of treatments, including chemotherapies and targeted therapies. We assume that tumor death produces most of the volatile signature,  breath sampling being done 3 weeks from the most recent cycle, i.e. away from the direct effect of the drug.

Focusing on subgroup analysis, we managed to distinguish SCLC from NSCLC with 100% sensitivity and 94% specificity (95% accuracy); adenocarcinoma from squamous cell with 75% sensitivity and 95% specificity (92% accuracy); mutation vs. non-mutation with 90% sensitivity and 50% specificity (81% accuracy). These are interesting and valid results; however, they do not reflect the need of sub-division of the results in assessing the use of breath sampling in monitoring response to anti-cancer treatment. Due to small sample size, we unfortunately were unable to draw a surrogate model for every subgroup.

In summary, standardization of assessment in tumor responsiveness is important in allowing effective comparisons between results of different cancer treatments. As strongly rooted in the clinical 'language', we expect RECIST to remain the primary criteria for the assessment of response to treatment in cancer patients, with advanced and novel monitoring techniques being used as adjuncts. This study offers a proof-of-concept of the idea that exhaled breath can provide a new modality in monitoring the response to anti-cancer treatment in patients with advanced lung cancer. It can help to define disease progression during conventional chemotherapy, molecular targeted therapy, or both. Monitoring tumor response to treatment during shorter intervals than currently available by CT scans should allow earlier decision-making regarding the chosen treatment. We believe that future research of exhaled breath analysis will yield excellent results that pave the way to using VOCs as markers for cancer diseases and for monitoring disease progression. Ideally, further studies require a larger study group and closer follow-up, which will give a more precise categorization of breath samples and statistically stronger results.

**ACKNOWLEDGEMENTS:**

This research project received funding from the FP7-Health Program under the LCAOS grant agreement (no. 258868). It was also supported by the Israel Cancer Association.

**REFERENCES:**

1. Bruzzi, J.F., et al., *Short-term restaging of patients with non-small cell lung cancer receiving chemotherapy.* J. Thorac. Oncol., 2006. **1**(5): p. 425-429.
2. Pujol, J.L., et al., *Lung cancer chemotherapy. Response-survival relationship depends on the method of chest tumor response evaluation.* Am. J. Respir. Crit. Care Med., 1996. **153**(1): p. 243-249.
3. Werner-Wasik, M., et al., *Assessment of lung cancer response after nonoperative therapy: tumor diameter, bidimensional product, and volume. A serial CT scan-based study.* Int. J. Radiat. Oncol. Biol. Phys., 2001. **51**(1): p. 56-61.
4. Therasse, P., et al., *New guidelines to evaluate the response to treatment in solid tumors.* J. Natl. Cancer Inst., 2000. **92**(3): p. 205-216.
5. Tisch, U. and H. Haick, *Nanomaterials for cross-reactive sensor arrays.* MRS bulletin, 2010. **35**(10): p. 797-803.
6. Tisch, U. and H. Haick, *Arrays of chemisensitive monolayer-capped metallic nanoparticles for diagnostic breath testing.* Rev. Chem. Eng., 2010. **26**(5-6): p. 171-179.
7. Amann, A., et al., *Lung cancer biomarkers in exhaled breath.* Expert Rev. Mol. Diagn., 2011. **11**(2): p. 207-217.
8. Amann, A., et al., *Analysis of exhaled breath for screening of lung cancer patients.* MEMO, 2010. **3**(3): p. 106-112.
9. Hakim, M., et al., *Diagnosis of head-and-neck cancer from exhaled breath.* BJC, 2011. **104**(10): p. 1649-1655.
10. Peng, G., et al., *Detection of lung, breast, colorectal, and prostate cancers from exhaled breath using a single array of nanosensors.* Br. J. Cancer, 2010. **103**(4): p. 542-551.
11. Peng, G., et al., *Diagnosing lung cancer in exhaled breath using gold nanoparticles.* Nat. Nanotechnol., 2009. **4**(10): p. 669-673.
12. Phillips, M., et al., *Detection of lung cancer with volatile markers in the breath.* Chest, 2003. **123**(6): p. 2115-2123.
13. Phillips, M., et al., *Volatile organic compounds in breath as markers of lung cancer: a cross-sectional study.* Lancet, 1999. **353**(9168): p. 1930-1933.
14. Poli, D., et al., *Determination of aldehydes in exhaled breath of patients with lung cancer by means of on-fiber-derivatisation SPME–GC/MS.* J. Chromatogr. B., 2010. **878**(27): p. 2643-2651.
15. van de Kant, K.D.G., et al., *Clinical use of exhaled volatile organic compounds in pulmonary diseases: a systematic review.* Respir. Res., 2012. **13**(177): p. 1-23.
16. Peled, N., et al., *Non-invasive breath analysis of pulmonary nodules.* J. Thorac. Oncol., 2012. **7**(10): p. 1528.
17. Song, G., et al., *Quantitative breath analysis of volatile organic compounds of lung cancer patients.* Lung Cancer, 2010. **67**(2): p. 227-231.
18. Montuschi, P., et al., *Diagnostic performance of an electronic nose, fractional exhaled nitric oxide, and lung function testing in asthma.* Chest, 2010. **137**(4): p. 790-796.
19. Dragonieri, S., et al., *An electronic nose in the discrimination of patients with asthma and controls.* J. Allergy Clin. Immunol., 2007. **120**(4): p. 856-862.
20. Fens, N., et al., *Exhaled breath profiling enables discrimination of chronic obstructive pulmonary disease and asthma.* Am. J. Respir. Crit. Care Med., 2009. **180**(11): p. 1076-1082.
21. Fens, N., et al., *External validation of exhaled breath profiling using an electronic nose in the discrimination of asthma with fixed airways obstruction*

|     | |
| --- | --- |
| | *and chronic obstructive pulmonary disease.* Clin. Exp. Allergy, 2011. **41**(10): p. 1371-1378. |
| 22. | Ligor, M., et al., *Determination of volatile organic compounds in exhaled breath of patients with lung cancer using solid phase microextraction and gas chromatography mass spectrometry.* Clin. Chem. Lab. Med., 2009. **47**(5): p. 550-560. |
| 23. | Peng, G., E. Trock, and H. Haick, *Detecting simulated patterns of lung cancer biomarkers by random network of single-walled carbon nanotubes coated with nonpolymeric organic materials.* Nano Lett., 2008. **8**(11): p. 3631-3635. |
| 24. | Dragonieri, S., et al., *An electronic nose in the discrimination of patients with non-small cell lung cancer and COPD.* Lung Cancer, 2009. **64**(2): p. 166-170. |
| 25. | Machado, R.F., et al., *Detection of lung cancer by sensor array analyses of exhaled breath.* Am. J. Respir. Crit. Care Med., 2005. **171**(11): p. 1286-1291. |
| 26. | Mazzone, P.J., et al., *Diagnosis of lung cancer by the analysis of exhaled breath with a colorimetric sensor array.* Thorax, 2007. **62**(7): p. 565-568. |
| 27. | Di Natale, C., et al., *Lung cancer identification by the analysis of breath by means of an array of non-selective gas sensors.* Biosens. Bioelectron., 2003. **18**(10): p. 1209-18. |
| 28. | Barash, O., et al., *Classification of lung cancer histology by gold nanoparticle sensors.* Nanomedicine, 2012. **8**(5): p. 580-589. |
| 29. | Broza, Y.Y., et al., *A nanomaterial-based breath test for short-term follow-up after lung tumor resection.* Nanomedicine, 2013. **9**(1): p. 15-21. |
| 30. | Zilberman, Y., et al., *Nanoarray of polycyclic aromatic hydrocarbons and carbon nanotubes for accurate and predictive detection in real-world environmental humidity.* ACS NANO 2011. **5**(8): p. 6743-6753. |
| 31. | Zilberman, Y., et al., *Carbon Nanotube/Hexa-peri-hexabenzocoronene Bilayers for Discrimination Between Nonpolar Volatile Organic Compounds of Cancer and Humid Atmospheres.* Adv. Mater., 2010. **22**(38): p. 4317-4320. |
| 32. | Konvalina, G. and H. Haick, *Effect of humidity on nanoparticle-based chemiresistors: a comparison between synthetic and real-world samples.* ACS Appl. Mater. Interfaces, 2011. **4**(1): p. 317-325. |
| 33. | Zilberman, Y., et al., *Spongelike structures of hexa-peri-hexabenzocoronene derivatives enhance the sensitivity of chemiresistive carbon nanotubes to nonpolar volatile organic compounds of cancer.* Langmuir, 2009. **25**(9): p. 5411-5416. |
| 34. | Brereton, R.G., *Chemometrics: Applications of mathematics and statistics to laboratory systems.* 1990, Ellis Horwood: Chichester. |
| 35. | Nishino, M., et al., *State of the art: response assessment in lung cancer in the era of genomic medicine.* Radiology, 2014. **271**(1): p. 6-27. |
| 36. | Organization, W.H., *WHO handbook for reporting results of cancer treatment.* 1979. |
| 37. | Nisman, B., et al., *Serum Thymidine Kinase 1 Activity in the Prognosis and Monitoring of Chemotherapy in Lung Cancer Patients: A Brief Report.* J. Thorac. Oncol., 2014. **9**(10): p. 1568-1572. |
| 38. | Nakhleh, M.K., Y.Y. Broza, and H. Haick, *Monolayer-capped gold nanoparticles for disease detection from breath.* Nanomedicine, 2014. **9**(13): p. 1991-2002. |
| 39. | Broza, Y.Y. and H. Haick, *Nanomaterial-based sensors for detection of disease by volatile organic compounds.* Nanomedicine, 2013. **8**(5): p. 785-806. |
| 40. | Hakim, M., et al., *Volatile organic compounds of lung cancer and possible biochemical pathways.* Chem. Rev., 2012. **112**(11): p. 5949-5966. |
| 41. | IARC, *Monographs on the Evaluation of Carcinogenic Risks to Humans: Schistosomes, Liver Flukes and Helicobacter Pylori.* 1994, Lyon: IARCPress. 233-246. |

42. Halliwell, B., J.M. Gutteridge, and C.E. Cross, *Free radicals, antioxidants, and human disease: where are we now?* The Journal of laboratory and clinical medicine, 1992. **119**(6): p. 598.

**LEGENDS:**

Figure 1: Graphic representation of breath sampling and analysis procedure. (A) collection of the breath sample in a single-step process into a chemically inert Mylar bag; (B) Transfer of the bag content to Tenax® sorbent tube; (C) Thermal Desorption to release the VOCs from the sorbent tube; (D) GC-MS to detect and analyze the compounds; (E) Sensor array containing Gold Nano Particles (GNPs) with different modifications and Poly Aromatic Hydrocarbons (PAH).

Figure 2: Hot-plot for the follow-up of LC states. Patients are represented by the rows and columns give successive breath samples for each patient. The scale represents the normalized resistance of the sensor, as described before, and units are given in ohms. Samples not identified by the sensor are marked by black blocks. The red tones in the color scale represent increased resistance of the sensor, indicating a worsening of the disease, whereas the green tones represent a decrease in resistance of the sensor and indicate a positive response to treatment.

**Table 1: Summary of personal clinical data (N=39)**

| Gender | Male | 31 (80%) | | |
|---|---|---|---|---|
| | Female | 8 (20%) | | |
| Age (years) | Range | 36-84 | | |
| | mean (SD) | 62 (±10.7) | | |
| BMI (kg/m²) | Mean (SD) | 24.5 (±4.3) | | |
| Smoking status | Never | 5 (13%) | | |
| | Previous | 27 (69%) | PY: 52 (±24) | |
| | Current | 7 (18%) | PY: 86 (±54) | |
| Histology | Non-small-cell lung carcinoma | 33 (85%) | Adenocarcinoma | 24 (73%) |
| | | | Squamous carcinoma | 4 (12%) |
| | | | Not Otherwise Specified | 5 (15%) |
| | small-cell lung carcinoma | 6 (15%) | All Extensive SCLC | |
| EGFR mutation | 7 (18%) | | | |
| ALK fusion | 1 (2.5%) | | | |
| Stage | 3A | 6 (15%) | | |
| | 3B | 3 (8%) | | |
| | 4 | 30 (77%) | | |

**Table 2: Categorization of breath samples**

| Group | Number of events (& breath samples) after one cycle of therapy | Total number of follow up events (& breath samples) |
|---|---|---|
| **Baseline sample** | | 39 |
| **Partial response (PR)** | 17 | 47 |
| **Stable disease (SD)** | 10 | 38 |
| **Progressive disease (PD)** | 5 | 11 |
| **TOTAL no. of breath samples** | | 135 |

**Table 3: VOCs from exhaled breath samples that discriminated significantly (p < 0.05) between the Baseline and DC groups and between DC and PD groups.**

| Retention time[min] | m/z | Chemical group | CAS no. | VOC name | BL (abundance + stderr) | PD (abundance+ stderr) | PR (abundance + stderr) | SD (abundance + stderr) | P-values (Wilcoxon non-parametric method) Baseline vs. DC | P-values (Wilcoxon non-parametric method) PD vs. DC |
|---|---|---|---|---|---|---|---|---|---|---|
| **25.6** | 93 | Alicyclic Hydrocarbons | 99-83-2 | alpha.-Phellandrene | 22043.8±3529.8 | 9960±2098.4 | 13673±1975.2 | 18206±3001.5 | 0.0075* ↓[a] | - |
| **20.3** | 104 | Aromatic compound | 100-42-5 | Styrene | 372346.1±59623.1 | 425657.8±91750.3 | 241379.8±34337.2 | 247320.8±41220.1 | 0.0241* ↓ | 0.0263* ↓ |
| **29.2** | 71 | Branched Alkanes | 6117-97-1 | Dodecane, 4-methyl | 17166.3±2748.8 | 7791.3±1318.4 | 11883.23±1234.11 | 13685.5±2280.917 | 0.0105* ↓ | - |

[a] *In breath analysis: a lower concentration of a VOC appears in the breath of DC samples compared to BL samples.*
[b] *In breath analysis: a lower concentration of a VOC appears in the breath of DC samples, compared to PD samples.*

**Table 4: Statistical classification success, using DFA and cross validation**

|  | BL vs. PR+SD | PR+SD vs. PD |
|---|---|---|
|  | PR+SD=positive<br>BL = negative | PD= positive<br>PR+SD = negative |
|  | Sensor set **1** | Sensor set **2** |
| **True positive** | 25 | 3 |
| **True negative** | 23 | 83 |
| **False positive** | 4 | 0 |
| **False negative** | 2 | 8 |
| **Sensitivity [a] (%)** | 93 | 28 |
| **Specificity [b] (%)** | 85 | 100 |
| **Accuracy [c] (%)** | 89 | 92 |
| **PPV (%)** | 86 | 100 |
| **NPV (%)** | 92 | 91 |

[a]Sensitivity=TP/(TP+FN).
[b]Specificity=TN/(TN+FP).
[c]Accuracy=TP+TN/(TP+TN+FN+FP).
[d]PPV - Positive predictive value =TP/(TP+FP).
[e]NPV - Negative predictive value = TN/( TN+FN).

**Table 5: Selected example of consecutive CT scans for patient #17 in the study (73 years old; SCLC), with the corresponding breath samples and NA-NOSE sensors' resistance (given in ohm).**

| CT scan | 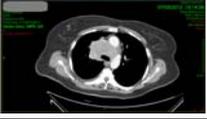 | 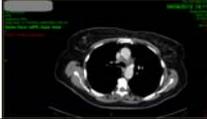 | 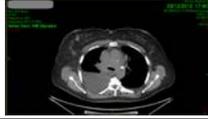 | 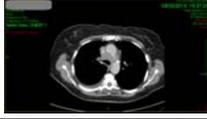 | 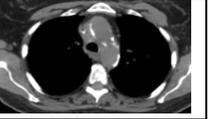 |
|---|---|---|---|---|---|
| Date of CT | 07/03/2012 | 08/08/2012 | 23/12/2012 | 03/02/2013 | 24/03/2013 |
| Date of breath sample | 05/04/2012 | 16/08/2012 | 21/11/2012 | 07/03/2013 | 02/04/2013 |
| Response to treatment | Baseline sample* | PR | PD | PR | SD |
| NA-NOSE (sensor's resistance [ohm]) | 1.252173 | 0.861627 | 2.761164 | 1.069338 | 0.788115 |

*before initiation of treatment

PD- progressive disease, PR- partial response, SD- stable disease